\documentclass{aastex}			
\usepackage{spr-astr-addons}	
\usepackage{url}

\usepackage{amsmath}			
\usepackage{physics}			
\usepackage{xcolor}				
\usepackage{MnSymbol}			
\usepackage{tabularx}			

\newcommand{\EM}[1]{\hspace{1pt}\textsc{e}\hspace{0pt}-\hspace{-1pt}#1 \phantom{.}}
\newcommand{\EP}[1]{\hspace{1pt}\textsc{e}\hspace{0pt}+\hspace{-1pt}#1 \phantom{.}}
\newcommand{\kB}{k_{\text{B}}}				
\newcommand{\GN}{G} 
\renewcommand{\ge}{g_{\star \epsilon}}		
\newcommand{\gn}{g_{\star n}}				

\begin{document}
\title{Entropy Production in a Lepton-Photon Universe}

\shorttitle{Entropy Production in a Lepton-Photon Universe}
\shortauthors{Lars Husdal \& Iver H. Brevik}

\author{Lars Husdal$^{1,a}$} \and \author{Iver Brevik$^{2,b}$}
\altaffiltext{1}{lars.husdal@ntnu.no}
\altaffiltext{2}{iver.h.brevik@ntnu.no}
\affil{$^a$Department of Physics, Norwegian University of Science and Technology}
\affil{$^b$Department of Energy and Process Engineering, Norwegian University of Science and Technology}
\email{lars.husdal@ntnu.no}
\email{iver.h.brevik@ntnu.no}

\abstract{
We look at the entropy production during the lepton era in the early universe by using a model where we exclude all particles except the leptons and photons. We assume a temperature dependent viscosity as calculated recently by one of us \citep{Husdal:2016Viscosity} with the use of relativistic kinetic theory. We consider only the bulk viscosity, the shear viscosity being omitted because of spatial isotropy. The rate of entropy production is highest just before the neutrinos decouple. Our results show that the increase in entropy during the lepton era is quite small, about 0.071\% at a decoupling temperature of $T=10^{10}~\mathrm{K}$. This result is slightly smaller than that obtained earlier by \citet{Caderni:1977}. After the neutrino decoupling, when the Universe has entered the photon era, kinetic theory arguments no longer support the appearance of a bulk viscosity. At high temperatures and a stable particle ratio, entropy production ($\dv*{\sigma}{T}$) goes as $T^{-8}$, with the total entropy ($\Delta \sigma$) increasing as $T^{-7}$. These rates go slightly down just before the neutrinos decouple, where $\Delta \sigma \propto T^{-6.2}$.}

\keywords{Viscous cosmology, bulk viscosity, lepton era, relativistic kinetic theory, entropy, entropy production}

\section{Introduction}
\label{sec:Introduction}
In an isotropic adiabatic expanding universe the entropy remains constant. However, for periods where there is a mixture of relativistic and non-relativistic particles, bulk viscous effects will arise, leading to an increase of the entropy. The magnitude of these depends strongly on the mean free paths of the particles. For the early universe, the bulk viscosity and resulting entropy production is at their highest just before the neutrinos decouple.

Back in 1977, Caderni and Fabbri calculated the entropy production to be 0.11\% during the lepton era \citep{Caderni:1977}. Nine years later, in 1986, Hoogeveen et al., used a more rigorous theory to calculate the viscosity, and obtained a result about 4.5 times larger \citep{Hoogeveen:1986}. This work was based on the relativistic kinetic theory developed by, among others, \citet{deGroot:RelativisticKineticTheory}, \citet{vanErkelens:1978}, \citet{vanErkelens:RelativisticBoltzmannTheoryV}, and \citet{vanLeeuwen:1986}.

In a previous paper \citep{Husdal:2016Viscosity} one of us used the paper by \citeauthor{Hoogeveen:1986} as a reference and expanded their model to include all three of the charged lepton pairs. This was done in a hypothetical universe where all particles except the leptons and photons were omitted. This allowed us to perform an isolated study of the viscous phenomena at temperatures where, strictly speaking, other particles would interfere. 

In this paper we use the theory by \citet{Brevik:1994} to calculate the entropy production for the same scenario, in a lepton-photon universe. We include all the charged leptons and look at entropy production caused by each of the three species.

\section{Setup and assumptions}
\label{sec:SetupAndConstraints}

\subsection{Definition of lepton era}
\label{sec:DefinitionOfLeptonEra}
The lepton era is normally defined to be between $T=10^{12}~\mathrm{K}$ and $T=10^{10}~\mathrm{K}$ ($\kB T = 100~\mathrm{MeV}$ to $\kB T = 1~\mathrm{MeV}$). At these temperatures the Universe is populated by photons, neutrinos, electrons and their antiparticles, and at the warmer end of the period, also some muons and pions. At approximately $T=10^{10}~\text{K}$ the neutrinos decouple and for lower temperatures we enter the photon era. For $T > 10^{12}~\mathrm{K}$ the appearance of baryons will increase rapidly, and at roughly $\kB T = 170~\mathrm{MeV}$ \citep{Kapusta:QGP} there is a phase transition to a quark-gluon plasma. At this time the energy content of the Universe is dominated by quarks and is hence called the quark era. In this paper, we disregard all particles except leptons and photons, and for our purposes, the lepton era is considered as the cosmic fluid before the neutrinos decouple.

\subsection{Assumptions in our model} 
\label{sec:Constraints}
The data for bulk viscosity are taken from \citet{Husdal:2016Viscosity}. We use the same assumptions, for the same reasons, as given in that paper. Since our study goes beyond the ``textbook'' lepton-era, our results for these high temperatures do not necessarily represent the real world. However, exploring this region in our modified universe gives us a better understanding of the entropy production with multiple particles. We make the following assumptions:

\begin{description}
\item[Pure lepton-photon universe:] 
We exclude all other particles to get a more isolated study of the entropy production at high temperatures. This leaves out any entropy production caused by the hadronic particles just after the quark-gluon plasma to hadron gas transition. At this time there would exist a large number of semi- and non-relativistic baryons and mesons, which we expect would create a significant viscous term. We recommend the paper by \cite{Dobado:2015} who discusses this era.

\item[No chemical potential:] 
The chemical potential for all particles is set to zero. This also excludes the asymmetry between matter and antimatter. This ratio is believed to be less than $10^{-9}$ \citep{Bennett:2013}, making it negligible for our study.

\item[No electromagnetic contribution to viscosity:]
The much larger cross sections for electromagnetic interactions compared to weak interactions mean that the weakly interacting neutrinos get a much larger mean free path, and resulting momentum transfer. We can disregard the viscous contribution from electromagnetic interactions for temperatures close up to $10^{14}~\mathrm{K}$ when the electromagnetic and weak forces approach each other in strength.

\item[No reheating from annihilation processes:] 
When the massive particles annihilate they should heat up the rest of the particles. This effect is not considered in our paper.

\item[Only elastic collisions:] 
We only consider elastic collisions in our calculations. The cross sections for inelastic collisions are of comparable sizes, and should thus not alter out results considerably \citep{Hoogeveen:1986}.

\item[Simple neutrino decoupling:] 
We use a simple cut-off for the neutrino interactions, which means they interact 100\% until a decoupling temperature and then 0\% thereafter. 

\end{description}

\section{Conditions in the early universe}
For a universe in thermal equilibrium (which is quite accurate for the early universe), the number density, $n$\footnote{The number density, $n$ is for all particles, not the net baryon number as in e.g. \cite{Weinberg:1971}.}, and energy density, $\epsilon=\rho c^2$, is given by the thermodynamic functions \cite{Husdal:2016DegreesOfFreedom}
\label{Eq:NEPS}
\begin{align}
  \text{Number density:}  \qquad  n
  &= \frac{\zeta(3)}{\pi^2} \gn(T) \frac{(\kB T)^3}{(\hbar c)^3} \;,
  \label{Eq:NumberDensityByG} \\
  \text{Energy density:}  \qquad\;\; \epsilon 
  &= \frac{\pi^2}{30} \ge(T)  \frac{(\kB T)^4}{(\hbar c)^3} \;,
  \label{Eq:EnergyDensityByG}
\end{align}
where $\kB$, $\hbar$, $c$, and $\zeta(3)$ are the Boltzmann constant, reduced Planck constant, speed of light, and the Riemann zeta function of value 3, respectively. $\gn$ and $\ge$ are the effective degrees of freedom for number density and energy density\footnote{$\ge$ is normally written without the $\epsilon$ subscript, as $g_\star$}. For high temperatures, when all particle species are created and destroyed at the same rate, we have $\gn = 15.5$ and $\ge = 17.25$. This is equal to the internal degrees of freedom for the particles, and is found by counting all the states of the particles, like spin, antiparticles, and including the distributional difference between fermions and bosons. This latter property results in fermions only contributing $3/4$ (number density) and $7/8$ (energy density) compared to bosons --- hence the difference in $\gn$ and $\ge$. The taus, muons, and electrons all contribute 3 to $\gn$ and 3.5 to $\ge$ at high temperatures, but as the temperature decreases and the annihilation rate becomes faster than the creation rate for these particles their effective contribution will decline. In a lepton-photon universe the evolution of $\gn$ and $\ge$ is given in Fig. \ref{Fig:GNGE}. In this and the following figures we use colored circles ($\medcircle$) in the graph, and triangles ($\medtriangleup$, $\medtriangledown$) on the axes, to mark the location where the rest mass of the taus, muons, and electrons equals that of the temperature ($mc^2 =\kB T$). We use magenta (taus), green (muons), and red (electrons) colors to represent the three different massive particle pairs.
\begin{figure}[h!]
	\includegraphics[width=\linewidth]{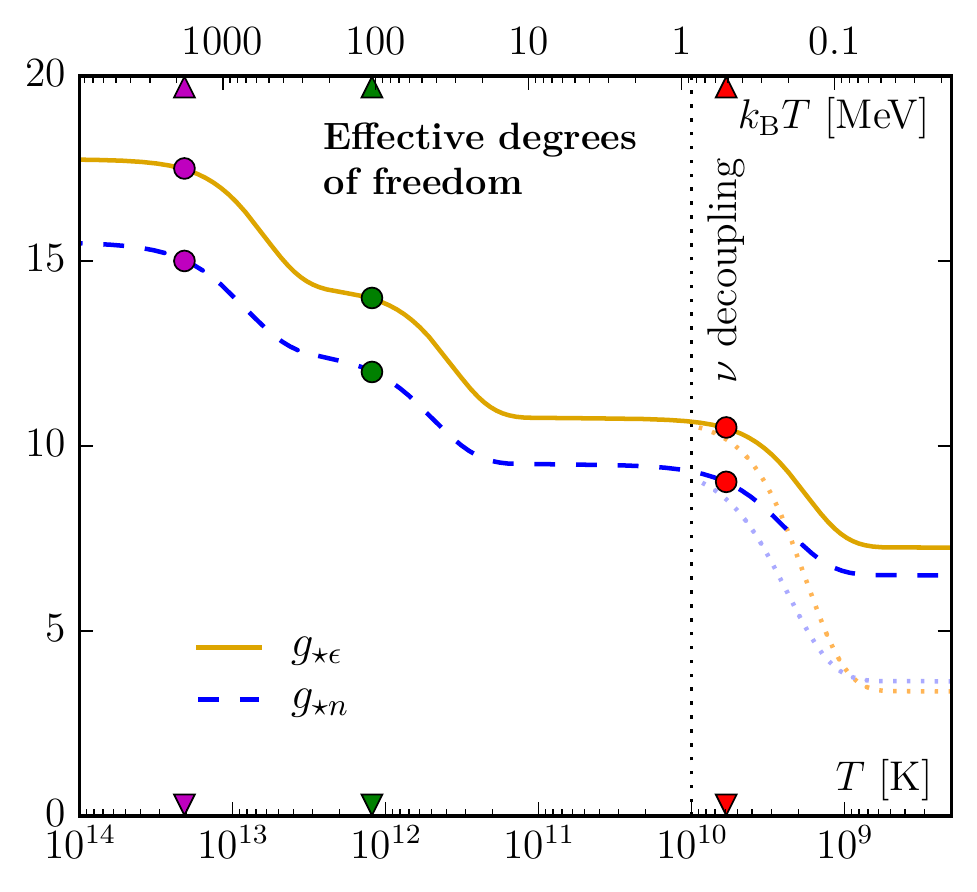}
	\caption{Effective degrees of freedom for energy density and number density. As the temperature decreases below the mass equivalent for the taus, muons and electrons, their contributions to $\gn$ and $\ge$ will drop, as we see as three dips in the two curves. The effective degrees of freedom from the neutrinos decrease after they decouple (as they are not heated by the decaying electrons and positrons). As we are interested in the difference in entropy production for slightly different decoupling temperatures, we keep the neutrino contribution to $\gn$ and $\ge$ constant in our calculations. The dotted curves show the standard representation of $\ge$ and $\gn$ where the neutrinos are not heated by the electron-positron annihilations, and is thus lower.  The masses of the three massive leptons are marked with magenta, green, and red circles ($\medcircle$) and triangles ($\medtriangleup$, $\medtriangledown$)}	
	\label{Fig:GNGE}
\end{figure}

Assuming that the early universe was flat and with negligible vacuum energy (cosmological constant equal to zero), we can write the first Friedmann equation as
\begin{equation}
  H^2 \equiv \left( \frac{\dot{a}}{a} \right)^2 = \frac{8\pi \GN}{3c^2} \epsilon \;,
  \label{Eq:Friedmann1}
  \end{equation}
where $\GN$ is the gravitational constant. We can use the energy density found in Eq. (\ref{Eq:EnergyDensityByG}), such that
\begin{equation}
  H^2 = \frac{8\pi \GN}{3c^2} \frac{\pi^2}{30} \ge(T)  \frac{(\kB T)^4}{(\hbar c)^3} \;.
  \label{Eq:FriedmannWithG}
  \end{equation}
For a lepton-photon universe $\ge$ will vary between $17.75$ and $3.36$, with a value of roughly $10.75$ at the time of the neutrino decoupling.

For a radiation dominated era, time can be calculated as a function of $\ge$ as follows \citep{Olive:2014PDG}: 
\begin{equation}
  t 	= \sqrt{\frac{90 \hbar^3 c^5}{32 \pi^3 \GN \ge}} (\kB T)^{-2} \;.
  \end{equation}

\section{The appearance of bulk viscosity}
\begin{figure}[h!]
	\includegraphics[width=\linewidth]{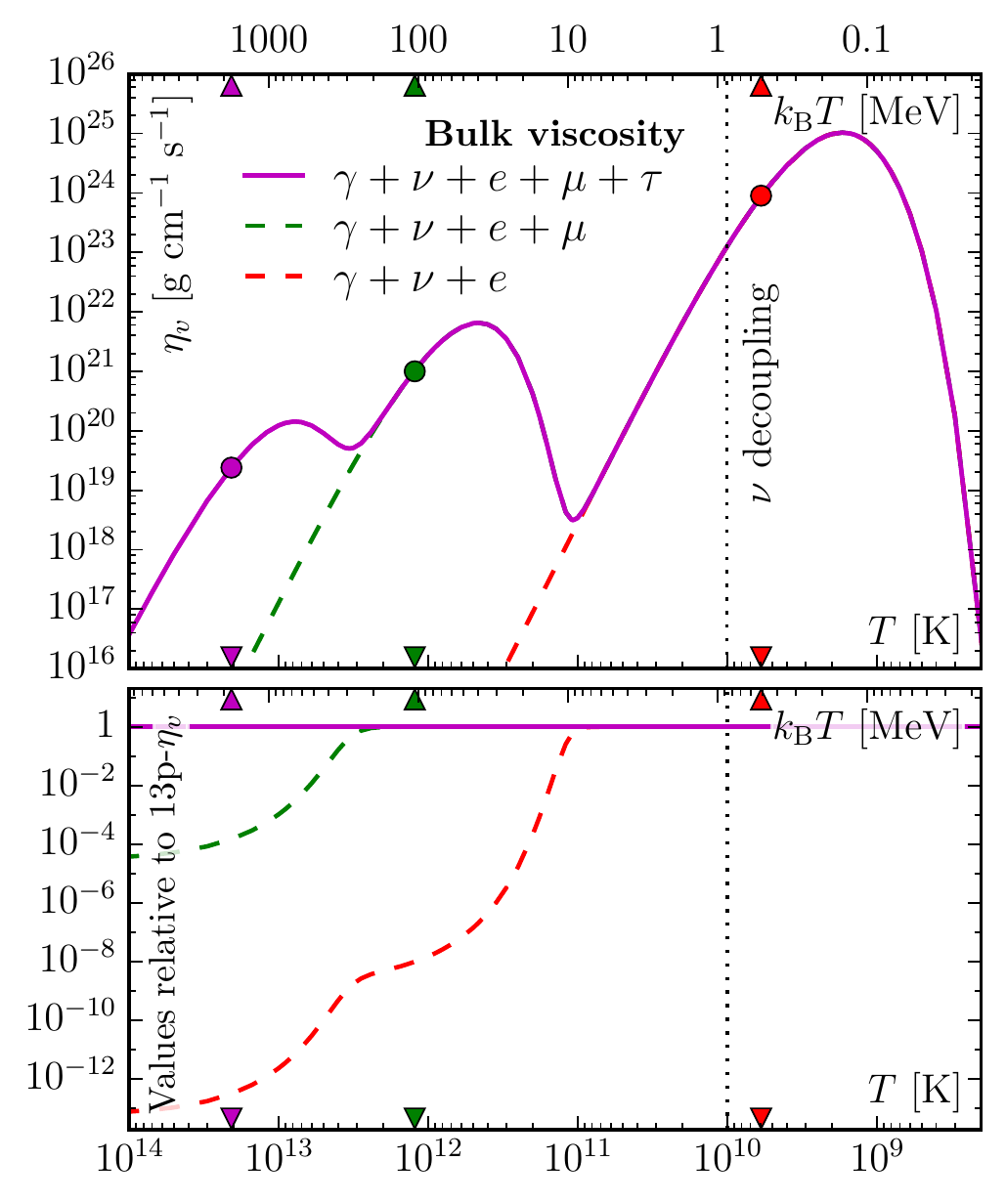}
	\caption{Bulk viscosity for models using two, four, or all six of the charged leptons in addition to the neutrinos and photon. The viscous effect is at its greatest when there is a large contribution of semi- and non-relativistic particles (which cool faster), and a long mean free path for the neutrinos (such that the temperature difference has time to build up). At $\kB T$ high above $m_{\tau, \mu, e}c^2$ the viscosity grows as $T^{-5}$. After reaching local maxima at around 1/3 of the rest mass of the particle in question, there will be an exponential drop. The viscosity at these local maxima goes roughly as $T^{-4/3}$ \citep{Husdal:2016Viscosity}}
	\label{Fig:BulkViscosity}
\end{figure}
Bulk viscous phenomena arise when two gas components expand differently. An expanding relativistic gas decreases in temperature as $a^{-1}$, while a non-relativistic gas decreases as $a^{-2}$. For semi-relativistic particles, we have something in between. Whenever we have a mixture at different temperatures there will be a heat transfer between the warmer components and the colder ones. The magnitude of this viscous factor depends on the temperature difference between the two components as well as the number ratio between them. The electromagnetic particles will interact frequently and never have time to build up a temperature difference, resulting in a negligible momentum transfer. The neutrinos, on the other hand, travel much farther before interacting and thus gain a larger temperature difference. The main component of the heat transfer is therefore due to the transfer of momentum from the neutrinos to the EM-coupled particles.

We emphasize that the number of particles in the mixture plays a big role. When one of the charged (massive) lepton species starts to drop quicker in temperature, it is almost instantaneously heated by all the remaining charged leptons and the photons. All the EM-coupled particles will thus collectively decrease in temperature a bit faster than the neutrinos. How fast depends on the number of EM-coupled particles. 
The decrease in temperature will though happen faster during $e^+e^-$ annihilations than during the $\mu^+\mu^-$, which again is faster than during $\tau^+\tau^-$ annihilations.

The bulk viscosity can be calculated as \citep{vanErkelens:1978, deGroot:RelativisticKineticTheory, Husdal:2016Viscosity}:
\begin{equation}
  \eta_v (T) = n \kB T \sum_k a_k \alpha_k \;,
  \label{Eq:BulkViscosity}
  \end{equation}
where $a_k$ are coefficients for particle $k$ for the linearized relativistic Boltzmann equations, describing the energy and momentum transfer, and $\alpha_k (n_k, T)$ describing the state of the system. $a_k$ and $\alpha_k$ are proportional to $T^{-7}$ and $T^{-2}$ for $\kB T \gg mc^2$, and as we see in Fig. \ref{Fig:BulkViscosity} the bulk viscosity ($\eta_v$) builds up as $T^{-5}$ as the massive leptons change to semi- and non relativistic velocities. However, as they eventually disappear we get an exponential decay of $\eta_v$.

\section{Entropy production}
To begin with, let us sketch the general relativistic formalism leading to the expression for the local rate of entropy production in the cosmic fluid. We assume a spatially flat FRW universe, and for simplicity, we put $c=1$ in this section where we are dealing with the general formalism. We include at first both the bulk viscosity $\eta_v$ and the shear viscosity $\eta_s$, but omit heat conduction. This kind of formalism can be found in various places, for instance in \cite{Weinberg:1971}. Here, we follow the formalism as given in \citet{Brevik:1994}.

If $U^\mu=(U^0,U^i)$ is the four-velocity of the fluid satisfying $U^\mu U_\mu=-1$, and if $h_{\mu\nu}=g_{\mu\nu}+U_\mu U_\nu$ is the projection tensor, we may write the energy-momentum tensor as
\begin{equation}
	T_{\mu\nu}=\rho U_\mu U_\nu+(p-3H\eta_v)h_{\mu\nu}-2\eta_s\sigma_{\mu\nu} \;.
\end{equation}
Here $H$ is the Hubble variable, and $\sigma_{\mu\nu}$ denotes the shear tensor,
\begin{equation}
	\sigma_{\mu\nu}=\theta_{\mu\nu}-Hh_{\mu\nu} \;,
\end{equation}
with
\begin{equation}
	\theta_{\mu\nu}=h_\mu^\alpha h_\nu^\beta U_{(\alpha; \beta)} 
\end{equation}
being the expansion tensor.

Let now $n$ denote the total number density of particles in the local rest frame, and let $\sigma$ be the nondimensional entropy per particle. The dimensional entropy per unit volume is thus $S=n \kB \sigma$.  The entropy current four-vector is
\begin{equation}
	S^\mu=n \kB\sigma U^\mu \;, 
\end{equation}
whose covariant divergence is
\begin{equation}
	{S^\mu}_{;\mu}=\frac{9\eta_v}{T}H^2+\frac{2\eta_s}{T}\sigma_{\mu\nu}\sigma^{\mu\nu} \;.
\end{equation}
The energy conservation for energy
\begin{equation}
	\dot{\rho}+3H(\rho+p)=9\eta_vH^2
\end{equation}
now gives, together with the conservation equation for particle number,
\begin{equation}
	(nU^\mu)_{; \mu}=0 \;,
\end{equation}
that $na^3=$ constant in the comoving frame. Setting $\eta_s=0$, this leads to
\begin{equation}
	\dot{\sigma}=\frac{9\eta_v}{n \kB T}H^2 \;,
\end{equation}
which is the desired result. This expression holds also in dimensional units, where $c$ is restored (i.e. the rest of our paper).

If we add the calculation for bulk viscosity done by \citet{Husdal:2016Viscosity} we find the following result for the total entropy production, $\Delta \sigma$:
\begin{align}
  \Delta \sigma	&= 
  \int_{t_0}^{t_1} \dot{\sigma} \dd{t}
  	= \int_{T_0}^{T_1} \dot{\sigma} \frac{\dd{t}}{\dd{T}} \dd{T} \notag \\
  &= \int_{T_0}^{T_1} \frac{9 \eta_v H^2}{n\kB T} \frac{\dd{t}}{\dd{T}} \dd{T} \;.
  \label{Eq:EntropyProduction}
\end{align}
Before we can calculate this integral we need to write the variables as temperature dependent. Using Friedmann's first equation (Eq. \ref{Eq:Friedmann1}) as well as number density and energy density from Eqs. (\ref{Eq:NumberDensityByG},\ref{Eq:EnergyDensityByG}) we get:  
\begin{align}
  \Delta \sigma	&= \frac{9}{\kB} \int_{T_0}^{T_1} 
  \dfrac{ \eta_v \dfrac{8 \pi \GN}{3c^2} \ge \dfrac{\pi^2}{30} \dfrac{(\kB T)^4}{(\hbar c)^3} }
  	    { \dfrac{\zeta(3)}{\pi^2} \gn \dfrac{(\kB T)^3}{(\hbar c)^3} T } \frac{\dd{t}}{\dd{T}} \dd{T} \\
		&= \frac{4 \pi^5 \GN}{5 \zeta(3) c^2}
		\int_{T_0}^{T_1} \eta_v(T) \frac{\ge(T)}{\gn(T)} \frac{\dd{t}}{\dd{T}} \dd{T} \;.
  \label{Eq:EntropyProduction2}
\end{align}

$\eta_v$, $\ge$, and $\gn$ are functions of temperature and need to be calculated numerically. Also, $\dv*{t}{T}$ needs to be calculated numerically.

\section{Results}
The main amount of entropy production happens just before the neutrinos decouple. The main factors of the entropy production come from the viscous term ($\eta_v$) and $\dv*{t}{T}$, which at high temperature ($\kB T \gg mc^2$) are proportional to $T^{-5}$ and $T^{-3}$. Disregarding neutrino decoupling leads to unphysical and extreme values for the entropy production. We include values at temperatures below this point in our plots, but clearly marked the neutrino decoupling, which, if not stated otherwise, is taken to occur at $10^{10}~\mathrm{K}$.

\subsection{Entropy production for all charged leptons}
\begin{figure}[h!]
	\centering
	\includegraphics[width=\columnwidth]{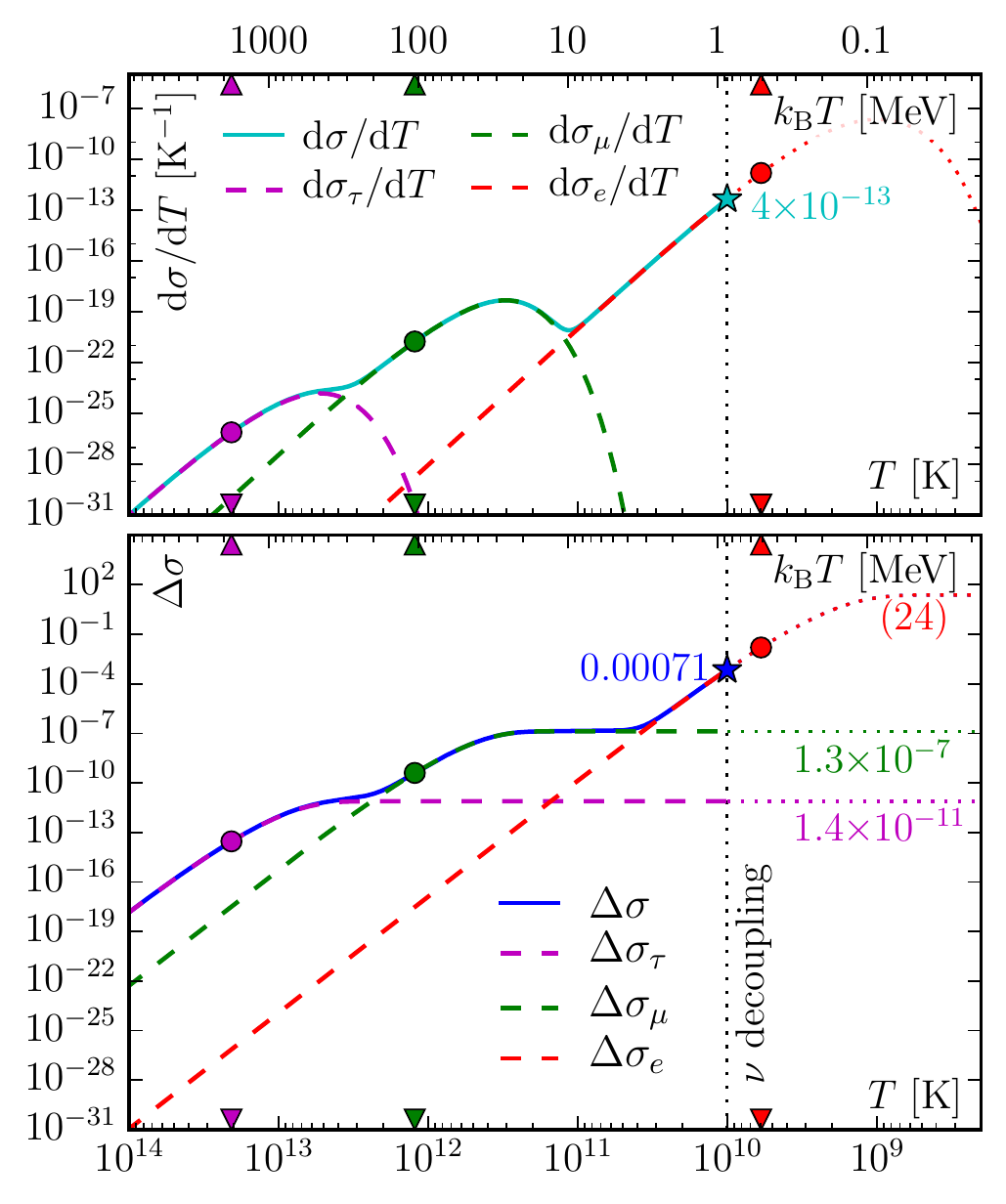}
	\caption{Upper plot: Entropy production as function of temperature ($\dv*{\sigma}{T}$). For high temperatures this grows as $T^{-8}$, before slowing down and reaching local maxima at around $\kB T=mc_x^2/3$. Lower plot: The total (accumulated) increase in entropy ($\Delta \sigma$), which grows as $T^{-7}$ for high temperatures. In both panels we have plotted the individual contributions by the $\tau^\pm$, $\mu^\pm$, and $e^\pm$ in dashed magenta, green and red colors. At a decoupling temperature of $10^{10}~\mathrm{K}$ the entropy production is $4.4 \times 10^{-11}~\mathrm{K^{-1}}$, resulting in a total increase of $7.1 \times 10^{-4} = 0.071\%$}
	\label{Fig:EntropyBySpecies}
\end{figure}

The main result is shown in Fig. \ref{Fig:EntropyBySpecies}. The upper plot shows the entropy production as a function of temperature, $\dv*{\sigma}{T}$. The lower plot shows the total (accumulated) entropy production, $\Delta \sigma$. We clearly see how the three different massive lepton-pairs dominate the entropy production at different times. The added curves (shown in dashed magenta, green, and red) show each of the charged leptons contributions to $\dv*{\sigma}{T}$ and $\Delta \sigma$. As seen in Table \ref{Tab:EntropyPerSpecie}, the final contribution of each lighter lepton pair outweighs those of the more massive ones. The final value of $\Delta \sigma$ depends on the neutrino decoupling temperature.
\begin{table}[h]
  \caption{Entropy production, $\dv*{\sigma}{T}$, and total increase in entropy, $\Delta \sigma$, caused by the different charged leptons. As the maximum for the electron-pair contribution is lower than the decoupling temperature, their actual contribution is that at $T = 10^{10}~\mathrm{K}$. The last row in our table is that for all particles at the decoupling (dc) temperature. In the first column the temperature is that of the local maxima for entropy production.}
  \begin{tabularx}{\columnwidth}{lccc}
  \hline\hline \\[-10pt]
					& Temp.
  					& \tt{max}(\hspace{-1pt}$\dv*{\sigma}{T}$\hspace{-1pt})
  					& Final $\Delta \sigma$  \\  
  					& $\mathrm{[K]}$	& $\mathrm{[K^{-1}]}$		 \\           				\cline{2-4} \\[-10pt]
	Taus			& $5.26\EP{12}$	& $1.52\EM{24}$	& $1.35\EM{11}$\\
	Muons			& $3.04\EP{11}$	& $4.68\EM{19}$	& $1.34\EM{7}$ \hspace{2pt}\\
	Electrons		& \textcolor{red}{$9.93\EP{8}$} \hspace{2pt}  
					& \textcolor{red}{$2.18\EM{8}$} \hspace{2pt} 
					& \textcolor{red}{$2.39\EP{1}$} \hspace{2pt} \\
	All at dc		& $1.00\EP{10}$	& $4.38\EM{13}$	& $7.10\EM{4}$ \hspace{2pt}\\
  \hline\hline
  \end{tabularx}
  \label{Tab:EntropyPerSpecie}
\end{table}

If we look more closely at $\dv*{\sigma}{T}$ in the upper panel of Fig. \ref{Fig:EntropyBySpecies} we see that the entropy production shares shape with the bulk viscosity plot in Fig. \ref{Fig:BulkViscosity}, but with an additional $T^{-3}$ factor. As expected, the entropy production has local maxima slightly to the colder end of the scale compared to the bulk viscosity. We find the local peaks for the tau and muon at temperatures roughly one-fourth that of their mass equivalent. For the hypothetical electron peak, this is roughly a sixth of its mass equivalent. The local peak values of $\dv*{\sigma}{T}$ increases roughly as $T^{-4.4}$ ($T^{-4.43}$ between the tau and muon peaks, and $T^{-4.29}$ between the muon and electron peaks.

\subsection{Entropy production at different decoupling temperatures}
\begin{figure}[h!]
	\centering
	\includegraphics[width=1\columnwidth]{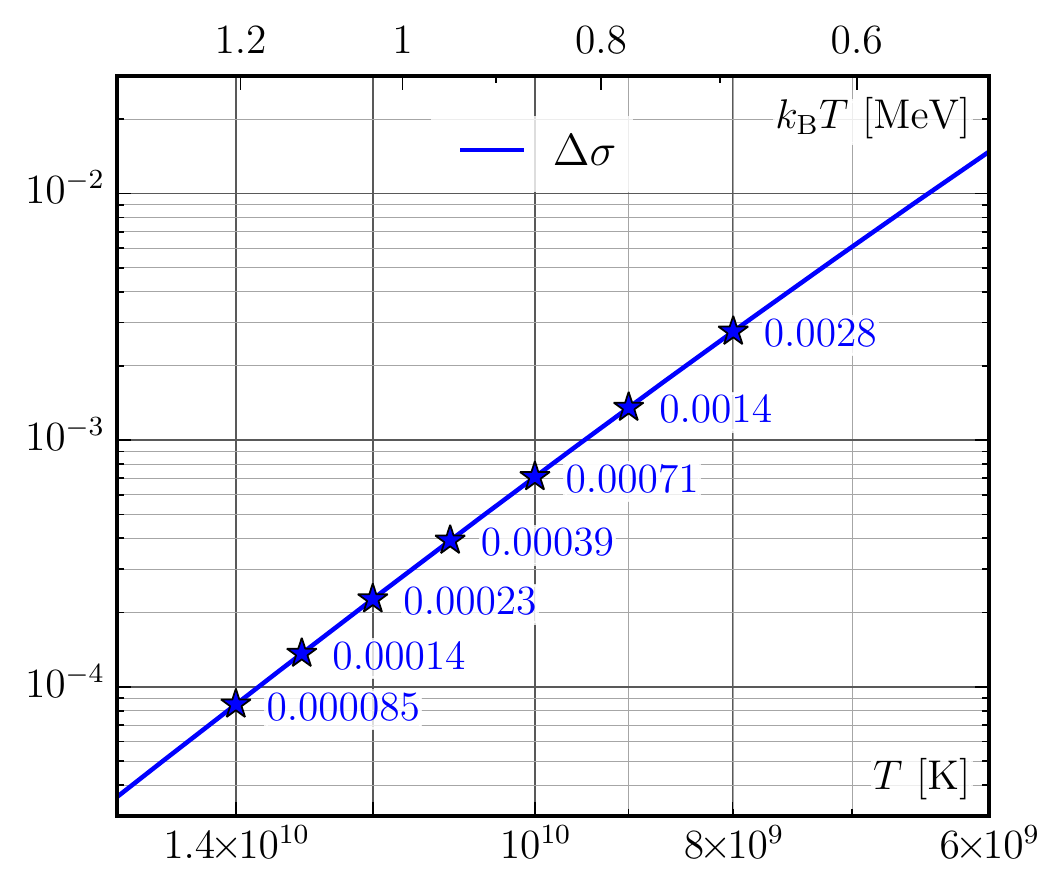}
	\caption{A closer look at the total increase in entropy ($\Delta \sigma$) zoomed in around the decoupling temperature. The total increase in entropy grows roughly as $T^{-6.2}$ around $T=10^{10}$ K. The final value of $\Delta \sigma$ is therefore very sensitive to the actual temperature at decoupling. $\Delta \sigma$ for different decoupling temperatures at intervals of 1 billion kelvins is marked with blue stars}	
	\label{Fig:EntropyAtDC}
\end{figure}

The total entropy production during the lepton era is extremely dependent on the temperature at the neutrino decoupling. If the neutrinos decouple at $T=10^{10}~\mathrm{K}$ we get an increase of entropy by about 0.071\%. Using a colder decoupling temperature of $\kB T = 1~\mathrm{MeV}$ ($T = 1.16 \times 10^{10}~\mathrm{K}$) we get a much smaller value, with $\Delta \sigma =0.017\%$. Fig. \ref{Fig:EntropyAtDC} shows the entropy production at different decoupling temperatures.

\subsection{Entropy production at high temperatures}
\begin{figure}[h!]
	\centering
	\includegraphics[width=1\columnwidth]{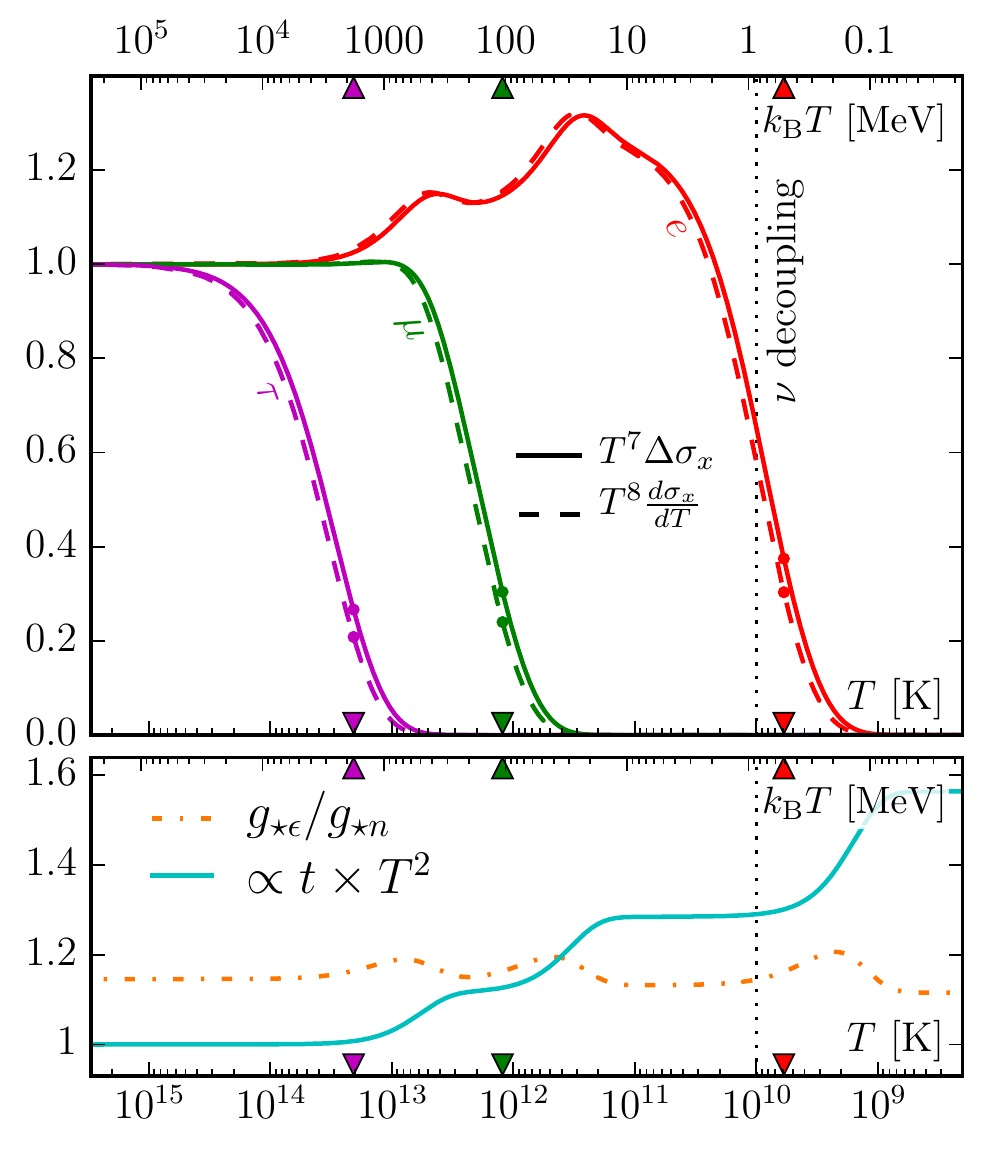}
	\caption{Upper panel: The rate of change for $T^8\dv*{\sigma}{T}$(shown in dashed lines) and $T^7\Delta \sigma$ (shown in solid lines) is constant for high temperatures, and are here normalized to one. We have here plotted the contributions due to the different charged leptons in magenta (taus), green (muons), and red (electrons). The fluctuations we see in the electron (and barely in the muon) case are due to changes in $\ge / \gn$ and the time-temperature relation during particle annihilations. During this process $\ge$ decreases and the temperature decreases slower with time. These two effects are shown specifically in the lower panel}
	\label{Fig:EntropyPower7}
\end{figure}

In Fig. \ref{Fig:EntropyPower7} we plot the evolution of the entropy production by temperature ($\dv*{\sigma}{T}$), and total entropy production ($\Delta \sigma$) due to the three charged lepton pairs multiplied by $T^8$ and $T^7$. For high temperatures where the particle ratio\footnote{i.e. when the creation rate and annihilation rate of the massive charged leptons are the same.} is essentially constant, the entropy production grows steadily as $T^{-8}$. We see some deviations of this trend for the neutrino-electron produced entropy (and slightly for the neutrino-muon equivalent). This has two reasons. Firstly, the temperature decreases slower when the effective degrees of freedom ($\ge$) decreases, and secondly the energy per particle first increases (because of the rest mass contribution of the massive particles) and then decreases (when the massive particles disappear). These two effects are shown in the lower plot in Fig. \ref{Fig:EntropyPower7}.

\subsection{Calculation method}
Our results are based on the viscosity calculated by \citet{Husdal:2016Viscosity}. These are calculated for 105 different temperatures, with higher densities at regions with higher particle annihilation rates, and around the neutrino decoupling temperature as this is when the entropy production is at its highest. These points are interpolated using Pythons SciPy package of the cubic kind. The number of steps for our numerical integration was crucial. Smaller steps equaled a smaller entropy production. Our experience was a convergence towards $\Delta\sigma=0.071\%$ with 1000 steps per decade. 10 steps per decade would give us $\Delta\sigma=0.30\%$. We used 100000 steps per decade from temperatures of $10^{16}~\mathrm{K}$ to $1.6 \times 10^{8}~\mathrm{K}$. 

\subsection{Difference from the Caderni and Fabbri Paper}
Our entropy is lower than that by \cite{Caderni:1977} by about two-thirds. This is uneasy because our viscous factor is 5 times as large. Using the viscosity as given by \cite{Caderni:1977} in our model for calculating entropy we get an end-result of 0.015\%, roughly 7 times smaller than their value of 0.11\%. Our numerical integration required a relatively high precision before converging to our given value. Using 10 steps per decade of temperature gave us an end result roughly five times larger than our high precision integration. If $\Delta \sigma$ would continue to grow as $T^{-7}$ (in the $\mathrm{e}^+\mathrm{e}^-$ case) its value would be roughly twice as large. Whether or not these arguments were considered by the \cite{Caderni:1977} unclear to us. However, we note that the sensitivity around the decoupling temperature requires a high precision numerical integration in this region. The differences from including the muon and tau particles is small (almost negligible) for the $\Delta \sigma$ value at the time of neutrino decoupling.

\section{Conclusions}
For high temperatures with a stable particle ratio, the entropy production ($\dv*{\sigma}{T}$) grows as $T^{-8}$, while the total increase in entropy ($\Delta\sigma$) grows as $T^{-7}$. At the time of neutrino decoupling $\Delta \sigma$ grows roughly as $T^{-6.2}$. The time of decoupling and how this happens is, therefore, essential to get a precise value for $\Delta \sigma$. Our model shows an increase in the entropy by $0.071\%$ at $T=10^{10}~\text{K}$, but this value will change a lot for small changes in the decoupling temperature. Using a decoupling temperature of $1~\mathrm{MeV}$ we get $\Delta \sigma = 0.027\%$.
Our result is two-thirds of that found by \citet{Caderni:1977}. This is a bit peculiar as our (and Hoogeveen et. al.'s) viscous term is roughly 5 times larger. For a more precise estimate of the entropy production during the lepton era, a rigorous theory of the decoupling should be used.

\acknowledgments
We would like to thank K\aa re Olaussen for inspiration and Eirik L\o haugen Fj\ae rbu for help with Python coding.


\bibliographystyle{spr-mp-nameyear-cnd}
\bibliography{Bibliography}

\end{document}